\documentclass[11pt,letterpaper]{article}
\usepackage{epsfig}
\usepackage[]{natbib}
\usepackage{amsmath,amsthm}
\usepackage{amssymb,amsfonts}
\usepackage{nccfoots,bbm,bm}
\usepackage{graphicx,color,caption}
\usepackage{booktabs,multirow,enumerate}
\usepackage{tikz,subfigure}
\usepackage{microtype}
\usepackage[colorlinks,citecolor=blue]{hyperref}
\usepackage[active]{srcltx}
\usepackage{xr}
\usepackage{amsmath}
\usepackage{setspace}
\usepackage{float}
\usepackage{fullpage,fancyvrb}
\usepackage{rotating}
\usepackage{dcolumn}
\usepackage{pdflscape}
\usepackage{rotating}
\usepackage{enumitem}

\usepackage[hmargin=1in,vmargin=1in]{geometry}
\newenvironment{packed_itemize}{
\begin{itemize}
  \setlength{\itemsep}{1pt}
  \setlength{\parskip}{0pt} 
  \setlength{\parsep}{0pt}
}{\end{itemize}}

\newcommand*{\defeq}{\mathrel{\vcenter{\baselineskip0.5ex\lineskiplimit0pt\hbox{\scriptsize.}\hbox{\scriptsize.}}}=}

\newcommand{\M}{\mathcal M}

\newcommand{\E}{{\mathbb{E}}}
\newcommand{\V}{{\mathbb{V}ar}}

\newcommand{\bone}{\mathbbm{1}}

\newcommand{\R}{{\mathbb{R}}}

\newcommand{\be}{\begin{equation}}
\newcommand{\ee}{\end{equation}}

\newtheorem{remark}{Remark}[section]
\numberwithin{equation}{section}
\theoremstyle{plain}

\def\qed{\rule{2mm}{2mm}}

\definecolor{myblue}{rgb}{0.2,0.4,1}
\definecolor{mygreen}{rgb}{0,0.5,0}

\begin{document}

\title{Hedging Forecast Combinations \\ With an Application to the Random
Forest}

\author{
Elliot  Beck\thanks{Second affiliation: Swiss National Bank,
8001 Zurich,  Switzerland.}\\
Department of Banking and Finance \\
University of Zurich\\
8032 Zurich, 
Switzerland\\
\href{mailto:elliott.beck@bf.uzh.ch}{elliot.beck@bf.uzh.ch} 
\and
Damian Kozbur\thanks{Corresponding author.}\\
Department of Economics\\
University of Zurich\\
8001 Zurich,
Switzerland\\
\href{mailto:damian.kozbur@econ.uzh.ch}{damian.kozbur@econ.uzh.ch} 
\and
Michael Wolf\\
Department of Economics \\
University of Zurich\\
8032 Zurich,
Switzerland\\
\href{mailto:michael.wolf@econ.uzh.ch}{michael.wolf@econ.uzh.ch} }

\date{August 2023}

\maketitle

\begin{abstract} 
This paper proposes a generic, high-level methodology for generating forecast combinations that would
deliver the optimal linearly combined forecast in terms of the mean-squared forecast
error if one had access to two population quantities: the mean vector
and the covariance matrix of the vector of individual forecast
errors. We point out that this problem is related to a mean-variance portfolio construction problem, in which portfolio weights correspond to forecast combination weights.  We allow negative forecast weights and interpret such weights as hedging over and under estimation risks across estimators.  This interpretation follows directly as an implication of the portfolio analogy.  We demonstrate our method's improved out-of-sample performance relative to standard methods in combining tree forecasts to form weighted random forests in 14 data sets.     
\end{abstract}

\begin{tabbing}
\noindent  
KEY WORDS: \=  Forecast combinations, nonlinear shrinkage, random forest.
\end{tabbing}

\noindent
JEL classification codes: C21, C53.

\newpage

\section{Introduction}

We visit the well-known and well-studied problem of
forecast combinations with a new angle. In this problem, one combines individual
forecasts of a univariate response variable using a vector-valued set
of regressors (or attributes) as input. The individual
forecasts are obtained by given number (or ensemble)
of forecasting methods and the combination is generally taken to be a
linear combination with weights summing up to one. Based on years of
hands-on experience, the consensus in the literature is that simple
averaging (or equal-weighting) of the individual forecasts is hard 
to~beat in practice.

This papers proposes a generic, high-level methodology that would
deliver the optimal linearly combined forecast in terms of the mean-squared forecast
error if one had access to two population quantities: the mean vector
and the covariance matrix of the vector of individual forecast
errors. We point out that this problem is related to the finance
problem of
portfolio selection, in which portfolio weights correspond
to forecast-combination weights.

In practice, the quantities needed above are unknown and must be
estimated based on a set of available (training) data. 
One of our
contributions is to suggest the use of nonlinear shrinkage in order to
estimate the covariance matrix as opposed to the standard 
(or canonical) choice in
the related literature, the sample covariance matrix. 

Another
contribution is that we allow for negative weights in the linear
combination of individual forecast, whereas the standard in the
literature is to enforce the weights to be non-negative. In order to
protect against weights that are unduly large in absolute value we borrow
an idea from the finance literature and enforce a ``gross-exposure
constraint'' on the weight vector, that is, an upper bound on its
$L_1$ norm (given by the sum of the absolute values of the
entries). In this way, we arrive at we call ``hedged forecast
combinations''.

As an application, we consider the random forest which is one of the
most popular and widely used supervised machine learning methods and
can be used for two main purposes: regression and classification. For
the purpose of regression, the random forest is a special case of an
equal-weighted forecast combination where the individual forecasting
methods are regression trees. We demonstrate empirically on a
collection of 14 benchmark
data sets that our methodology
applied to the random forest, called ``hedged random forest'',
improves the forecasting performance of the standard random forest,
especially for smaller training sets.

In the remainder of this paper, 
Section~\ref{sec:method} presents the general methodology, 
Section~\ref{sec:application} provides an application to the random
forest, and
the appendix contains various robustness checks.

\section{Methodology}
\label{sec:method}

\subsection{General Description}
\label{sec:high-level}

The goal is to forecast (or predict)\footnote{In this paper, the terms
  ``forecast'' and ``prediction'' are used interchangeably. 
  Arguably, some people associate with ``forecast'' a time series
  setting but we do not.}
a random variable $y \in \R$ based on a set of variables
(or attributes) $x \in \R^d$. Denote a generic forecast by $\hat
f$. Then its mean-squared error (MSE) is given~by
$$
\mbox{MSE}(\hat f) \defeq  \E \bigl (y - \hat f(x) \bigr)^2~.
$$
Here and below the moments are, of course, obtained under the joint
distribution governing the random vector $v \defeq (y, x')' \in
\R^{1+d}$ and assumed to exist.
Letting
$$
\mbox{Bias} (\hat f) \defeq \E \bigl (y - \hat f(x) \bigr )
\quad \mbox{and} \quad
\mbox{Var} (\hat f) \defeq \V \bigl (y - \hat f(x) \bigr )
\defeq \E \bigl ( (y - \hat f(x) \bigr )^2 -
\bigl [\E \bigl ( (y - \hat f(x) \bigr )\bigr ]^2~,
$$
there exists the well-known decomposition
\begin{equation} 
\mbox{MSE}(\hat f) = \mbox{Bias}^2(\hat f) + \mbox{Var}(\hat f)~.
\end{equation}
The oracle that minimizes the MSE is given by the
conditional expectation
\mbox{$\hat f_{\text{or}}(x) \defeq \E(y|x)$}
but is not available in practice; for example,
see \citet[Section 2.9]{hayashi:2000}.

This paper considers combinations of a given set of $p$ forecasting
methods (or forecasting models), denoted by $\{\M_j\}_{j=1}^p$. The number of methods,
$p$, is assumed to be exogenous and fixed. Although we do not make
this explicit in the notation, methods may be data-dependent in the
sense that, for example, certain parameters are fitted based on
observed data (such as regression parameters).

There exists an extensive literature on forecast combinations; for
example, see \citet[Chapter 14]{elliott:timmermann:2016},
\cite{wang:hyndman:li:Kang:2022}, and the references therein. The
consensus seems to be that simple averaging (or equal weighting), given by
$$
\hat f_{\text{AV}}(x) \defeq \frac{1}{p} \sum_{j=1}^p \M_j(x)~,
$$
is hard to beat by more general linear combinations of the kind
\begin{equation} \label{eq:rf-w}
\hat f_w(x) \defeq \sum_{j=1}^p w_j \M_j(x) \quad \mbox{with}
\quad w \defeq (w_1, \ldots, w_p)' \quad \mbox{and}
\quad \sum_{j=1}^p w_j= 1~.
\end{equation} 
Nevertheless, our aim is to find a method for selecting a set of
weights $w$
that does improve the (out-of-sample)
MSE of simple averaging, at least `on balance'.

Denote by $e_j \defeq y - \M_j(x)$ the forecast error made by model $M_j$
and collect these errors into the vector $e \defeq (e_1, \ldots,
e_p)'$ with expectation (vector) and covariance matrix
$$
\mu \defeq \E(e) \quad \mbox{and} \quad \Sigma \defeq \V(e)~.
$$
The MSE of the forecast \eqref{eq:rf-w} is then given by
$$
\mbox{MSE} (\hat f_w) = (w' \mu)^2 + w' \Sigma w~.
$$
Therefore, the optimal (in terms of the MSE) forecast in the class
\eqref{eq:rf-w} is the solution of the following optimization problem:
\begin{align}
& \min_w \; (w' \mu)^2 + w' \Sigma w\label{eq:m1} \\
\mbox{s.t.} \quad & w'\bone = 1~, \label{eq:m2}
\end{align}
where $\bone$ denotes a conformable vector of ones.
Problem \eqref{eq:m1}--\eqref{eq:m2} is a convex optimization
problem and can, in principle, be solved quickly with readily
available software, even for large dimensions~$p$.

The problem in practice is that the inputs $\mu$ and $\Sigma$ are
unknown. A feasible solution is to~replace them with sample-based
estimates $\hat \mu$ and $\hat \Sigma$, which is an application of the general
``plug-in method''.


Being agnostic, for the time being, abut
the nature of the estimators $\hat \mu$ and $\hat \Sigma$, we then
solve the feasible optimization problem
\begin{align}
& \min_w \; (w' \hat \mu)^2 + w' \hat \Sigma w\label{eq:mf1} \\
  \mbox{s.t.} \quad & w'\bone = 1 \quad \mbox{and} \label{eq:mf2}\\
   \quad & ||w||_1 \le \kappa~, \label{eq:mf3}                   
\end{align}
where 
$||w||_1 \defeq \sum_{j=1}^p |w_j|$ 
denotes the $L_1$ norm of
$w$ and $\kappa \in [1, \infty]$ is constant chosen by the user.
Assuming succinctly that the estimator $\hat \Sigma$ is symmetric and
positive semi-definite,
the optimization problem
\eqref{eq:mf1}--\eqref{eq:mf3} is still of convex nature and can
be solved easily and quickly in practice, even for large dimensions $p$.
We shall denote the solution to this optimization problem by $\hat w$.

The addition of the constraint \eqref{eq:mf3} is motivated by the
related problem of {\em portfolio selection} in~finance, in which
context the constraint is called a ``gross-exposure
constraint''. Adding this type of constraint to the infeasible problem
\eqref{eq:m1}--\eqref{eq:m2} clearly would result in a (weakly)
worse solution for any value $\kappa \in [1, \infty)$. But in the
feasible problem, which must use estimated instead of true inputs, the
constraint typically helps. The intuition here is that replacing $\mu$
and $\Sigma$ with respective
estimates $\hat \mu$ and $\hat \Sigma$ can lead to
unstable and underdiversified solutions that look good in sample
(or in the training set) but
perform badly out of sample, especially when the number \mbox{of models,
$p$,} is not (exceedingly) small relative to the sample size
relevant to the estimation of  $\mu$ and $\Sigma$;
for example, see \cite{jobson:korkie:1980},
\cite{michaud:1989}, and \cite{jagannathan:ma:2003}.

In the extreme case $\kappa = 1$, the
weights are forced to be non-negative, that is, $w_i \ge 0$, which is
called a ``no-short-sales constraint'' in finance. Imposing this
constraint is standard in the forecast-combination literature but it
might well lead to sup-optimal performance because of not giving
enough flexibility to the solution of the problem
\eqref{eq:mf1}--\eqref{eq:mf3},
At the other end of
the spectrum, the choice $\kappa = \infty$ corresponds to 
removing the constraint \eqref{eq:mf3}, which may also lead to
sub-optimal performance
for the reasons mentioned above. Staying away from either extreme,
there is ample evidence in the finance literature
that choosing $\kappa \in [1.5, 2.5]$
typically results in improved forecasting performance, and that the
exact choice in this interval is not overly critical;
for example, see \cite{demiguel:ms:2009}.

Because the constraint \eqref{eq:mf3} protects the user against extreme
``positions'', that is, against weights~$\hat w_j$ that are unduly large
in absolute value,
we call our approach ``hedging forecast
combinations''.\footnote{For example, according to Merriam-Webster
  (online version) the verb ``to hedge against'' means ``to protect
  oneself from (something)''.}

\subsection{Theory}

The solution to the convex optimization problem
\eqref{eq:mf1}--\eqref{eq:mf3} is continuous in the inputs $\hat \mu$
and $\hat \Sigma$.  Therefore, with the
choice $\kappa \defeq \infty$, its solution, denoted by $\hat w$ would
lead to an asymptotically optimal forecast \mbox{combination $\hat
f_{\hat w}$} based on consistent estimators $\hat \mu$ and $\hat
\Sigma$.  Stating this fact in a theorem is possible, but as this is a routine matter we find it outside the scope the basic research content of this paper. First, this fact has been recognized
before.
Furthermore, in practical application, the relevant property is the finite-sample performance
of the forecast $\hat f_{\hat w}$ and, so far, the evidence based on
simulation studies and empirical applications to real-life data sets
indicates that such forecast combinations, on balance, do not
outperform $\hat f_{\text {AV}}$, that is, simple averaging; again,
see \citet[Chapter 14]{elliott:timmermann:2016},
\cite{wang:hyndman:li:Kang:2022}, and the references therein.

Therefore, our goal is isolated to finding a forecast combination $\hat f_{\hat w}$ that,
on balance, outperforms $\hat f_{\text{AV}}$ in empirical applications
to commonly used benchmark data sets.\footnote{On the other hand, we shall
abstain from any simulation studies, since we could tilt the data
generating process arbitrarily to our favor.} 


Our preceding high-level description is agnostic about (i) the nature of
the forecasting methods $\{\M_j\}_{j=1}^p$ and (ii) 
the estimation of the mean (vector)
$\mu$ and the covariance matrix~$\Sigma$ of the corresponding vector
of forecast errors $e \in \R^p$. This estimation is crucial
to the performance of the proposed forecast-combination method.
We assume the existence of a data set $\{v_i\}_{i=1}^n$ with $v_i \defeq
(y_i, x_i')'$ and consider two cases.

When $\{v_i\}_{i=1}^n$ is an independent and identically distributed
(i.i.d.) sample where the distribution of $v_i$ is equal to the
distribution of $v$. In this case there exists a well-established
literature on how to generate pseudo-out-of-sample forecast errors,
with the most popular technique being cross-validation; for example,
see \citet[Chapter 12]{efron:hastie:2016} and
\citet[Chapter~7]{hastie:tibshirani:freedman:2017}. Another option is
to use in-sample errors, or residuals; this option has a bad
reputation because in-sample errors tend to be systematically smaller
(in magnitude) compared to (actual) out-of-sample errors due to the
well-known phenomenon of ``overfitting''. However, for our purposes
this may not be a serious problem; see Remark~\ref{rem:scale}
below. Whether one computes pseudo-out-of-sample or in-sample errors,
the results are collected in a matrix $R$ based on which one estimates
$\mu$ and~$\Sigma$.
To~this end, one can use sample counterparts (that is, sample mean
and sample covariance matrix), shrinkage methods, penalized estimation
schemes, etc; Having said this, we
shall restrict attention to shrinkage methods as an
alternative to sample counterparts.
For shrinkage
estimation of  a mean vector, see \cite{hansen:2016},
\cite{bodnar:okhrin:parolya:2019}, and the references therein; for
shrinkage estimation of a covariance matrix, see
\cite{ledoit:wolf:power} and the references therein.

Note that other data settings are also possible, for instance when $\{v_1, \ldots, v_n, v\}$ is a stationary time series. Also in
this case the recommendation is to base the estimation of $\mu$ and
$\Sigma$ on a collection of pseudo-out-of-sample or in-sample
forecast errors~$R$. How
to generate those is less well established compared to the i.i.d. case
but proposals do exist; for example, see \cite{bergmeir:benitez:2012} and
\cite{bergmeir:hyndman:koo:2018}. If one prefers shrinkage estimation
over sample counterparts, ideally one should used methods designed for
time-series data; for example, see \cite{sancetta:2008} and
\cite{engle:ledoit:wolf:2019} for shrinkage estimation of the
covariance matrix. Having said this, the case of stationary time
series is, arguably, more difficult in practice. Compared to an
i.i.d.\ sample it would take a larger sample size, generally, to have
a similar chance of outperforming $\hat f_{\text{AV}}$, that is,
simple averaging; but especially macroeconomic time series only have
relatively small sample sizes. Furthermore, many real-life time series
(even after detrending and deseasonalizing) may still not be
stationary because of structural breaks, for example.

\begin{remark}[Scale invariance]\label{rem:scale}
  \rm
The solution $\hat w$ to the optimization problem
\eqref{eq:mf1}--\eqref{eq:mf3} remains unchanged if $\hat \mu$ and
$\hat \Sigma$ are replaced by $c \hat \mu$ and $c^2 \hat \Sigma$,
respectively, for any constant $ c \in (0, \infty)$. Therefore, it is
not important that the estimators $\hat \mu$ and $\hat \Sigma$ get the
`levels' of the true quantities $\mu$ and $\Sigma$ right.
In
particular, the use of in-sample (or training-set) errors in the
construction of $\hat \mu$ and~$\hat \Sigma$ can still lead to
favorable performance of the forecast combination $\hat f_{\hat w}$
even if such errors are systematically smaller (in magnitude) compared
to out-of-sample errors because of in-sample (or training-set)
overfitting. Instead of approximating the actual entries of $\mu$ and
$\Sigma$, the corresponding estimators $\hat \mu$ and $\hat \Sigma$ only
need to approximate the entries relative to each other in order 
for~$\hat f_{\hat w}$ to~outperform $\hat f_{\text{AV}}$. That may still
not be a trivial task, but it is certainly an easier task.~\qed
\end{remark}

\section{Empirical Application: The Random Forest}
\label{sec:application}

\subsection{Background}

The random forest is one of the most popular tree-based methods in supervised
machine learning; at time of this writing the original paper
\cite{breiman:2001} has more than 110,000 Google Scholar citations
already.
If the variable $y$ is categorical, the random forest is used for
{\em classification}; if the variable $y$ is numerical, the random
forest is used for {\em regression}. In this application, we focus on regression
only.

A review of the mechanics of the random forest is as follows.  First, one grows an
ensemble of {\em de-correlated} trees; these correspond to the
forecasting methods $\{\M_j\}_{j=1}^p$ in our context.
Second, one uses the simple average of
the trees or, alternatively put, the equal-weighted ensemble. This
means the individual forecasts of the trees are simply averaged to
arrive at a final, combined forecast; this corresponds to the forecast
combination $\hat f_{\text{AV}}$ in our context.
For a textbook treatment on the random forest the
reader is referred to, for example, \citet[Chapter
15]{hastie:tibshirani:freedman:2017}.


In our analysis below we shall study whether it is possible to find a
forecast combination $\hat f_{\hat w}$ that can, on balance,
outperform the (standard) random forest $\hat f_{\text{AV}}$. To this
end we will use a number of benchmark data sets (some artificial, some
real) from the literature. Importantly, we shall restrict focus on
cross-sectional  (or i.i.d.) data sets.

\subsection{Data}

To assess the performance of our proposed method relative to the
standard (equal-weighted) random forest,
we use 14 
data sets sourced from the Penn Machine Learning Benchmarks (PMLB)
database; see \cite{romano2021pmlb}.
PMLB serves as a comprehensive
repository of benchmark data sets, designed specifically for the
evaluation and comparison of supervised machine learning
methods. Each data set is available on the website
\url{www.openml.org}, which also provides
access to metadata and descriptions.
In particular, we select all data sets whose number of observations
ranges from 6,000 to 100,000\footnote{After manually
  inspecting all data sets, we removed two data
  sets: 294\_satellite\_image was removed because it is falsely
  classified as a regression task and 503\_wind was removed because it
  contains time series data.}.
The lower bound of 6,000 was chosen to
ensure a test set of size at least 1,000 in all scenarios, as
will become apparent below.
Note that our selection
contains both artificial and real-world data
sets. For example, the data set 201\_pole is a real-world
telecommunication data set first used by \citet{weiss:indurkhya:1995}; on
the other hand, 564\_fried is an artificial data set 
introduced by \citet{friedman:1991} and also described in
\citet{breiman:1996}. Table~\ref{table:datasets} lists the data sets
used in this analysis together with the corresponding numbers of
observations and
numbers of attributes (or $x$-variables).

\begin{table}
\centering
\begin{tabular}{lrr}
\hline
Name   & \# Observations  & \; \; \# Attributes \\
\hline
197\_cpu\_act & 8,192 & 21  \\
201\_pol & 15,000 & 48  \\
215\_2dplanes & 40,768 & 10  \\
218\_house\_8L & 22,874 & 8 \\
225\_puma8NH & 8,192 & 8  \\
227\_cpu\_small & 8,192 & 12  \\
344\_mv & 40,768 & 10\\
537\_houses & 20,640 & 8\\
562\_cpu\_small & 8,192 & 12\\
564\_fried & 40,768 & 10\\
573\_cpu\_act & 8,192 & 21\\
574\_house\_16H & 22,784 & 16 \\
1193\_BNG\_lowbwt & 31,104 & 9 \\
1199\_BNG\_echoMonths & 17,496 & 9\\
\hline
\end{tabular}
\captionof{table}{Data sets used.}
\label{table:datasets}
\end{table}

\subsection{Implementation}

Throughout, we
consider a fixed range of training-set (sample) sizes
$200 \le n \le 5,000$.
For a given data set and a given training-set
size, we then draw $n$ observations at random (without
replacement) as the training set and the
remaining observations constitute the test set; as a consequence,
when $n=5,000$ we
need at least 6,000 observations in the original data set to~obtain a
test set of size 1,000.
We next train a random forest on the training set 
using the ``ranger library'' implemented in the
programming language  {\sc R},
where the various hyperparameter are set to the
defaults recommended by \citet{ranger:2017}; in~particular, the number
of trees is set to $p=500$ as per default.

After training the random forest, we extract the forecasts of each
tree on the entire training set and thus obtain a residual
(or in-sample error) 
matrix
of size $n \times p$.
We do not, for a given tree, extract predictions on the corresponding
out-of-bag observations only (that is, on the subset of the training set
not used in pruning the particular tree) because in this way we would
not obtain a full $n \times p$ matrix of residuals but instead a
matrix that would contain a large number of missing values.\footnote{On average,
there would be about $1 - 1/e \approx 63.2\%$ of missing values}
But as explained below, we need a
full matrix $R$ for the estimation of $\Sigma$, if not for the estimation
of $\mu$ necessarily.

The various inputs to the feasible optimization problem
\eqref{eq:mf1}--\eqref{eq:mf3} are chosen as follows.
First, for the estimation of $\mu$, we use the (column-wise) sample
mean of $R$; we
also experimented with some shrinkage estimators instead but the
results remained virtually unchanged.
Second, for the estimation of $\Sigma$, we apply nonlinear shrinkage to $R$; in
particular, we use the quadratic
inverse shrinkage (QIS) estimator of
\cite{ledoit:wolf:2022}.\footnote{Nonlinear shrinkage estimation of covariance matrices
  are reviewed in \cite{ledoit:wolf:power}.}
Note here that nonlinear shrinkage requires a full 
matrix $R$ as an input.
Third, for the gross-exposure constraint $\kappa$ in \eqref{eq:mf3}, we use
$\kappa \defeq 2$. Appendix~\ref{app:robust}
runs robustness checks that
consider (i) the sample covariance matrix based on~$R$, rather than
nonlinear shrinkage, as the estimator $\hat \Sigma$ and (ii) alternative
values of the gross-exposure constraint $\kappa$.

Thus all inputs to the
optimization problem \eqref{eq:mf1}--\eqref{eq:mf3} are now in place.
The solution
$\hat w$ assigns weight $\hat w_j$ to tree $\M_j$ rather than weight
$1/p$ as for the standard random forest (RF). We call this weighted random
forest the ``hedged random forest'' (HRF).

For each method, RF and HRF, fitted on the training~set we obtain a MSE on
the test~set, denoted by MSE$_{\text{RF}}$ and  MSE$_{\text{HRF}}$,
respectively. In order to eliminate, or at least mitigate, randomness
due to the random choice of $n$ observations as the test set, we then
repeat this process (independently) $B$ times. As the final
performance measure, we report the following
root-mean-squared-error (RMSE) ratio:

\begin{equation}\label{eq:rmse-rat}
\text{RMSE}_{\text{HRF}/\text{RF}} \defeq 
  \frac{
  \sqrt{\frac{1}{B} \sum_{b=1}^B  \text{MSE}_{\text{HRF,b}}}}
  {\sqrt{\frac{1}{B} \sum_{b=1}^B  \text{MSE}_{\text{RF,b}}}}~.
\end{equation}
This means that, for each method, we average the MSE values over the
$B$ repetitions and then take the root to arrive at individual RMSE
values. Finally, we take the ratio of the two RMSE values. Values
greater than one of this ratio speak in favor of RF whereas values
smaller than one speak in favor of HRF. Our results below are all
based on $B=100$ repetitions; larger values of $B$ leave the~results
virtually unchanged.

In this way, for any training-set size $n$, we get 14 RMSE
ratios~\eqref{eq:rmse-rat}, one for each data set listed in
Table~\ref{table:datasets}. We convert the 14 ratios into a
boxplot and then line up the boxplots for
$n \in \{200, 400, 600, 800, 1000, 2000, 3000, 4000, 5000\}$ in
Figure~\ref{fig:main}.

\medskip
\begin{center}
\captionsetup{type=figure}
\includegraphics[width=5.5in,height=5in]
{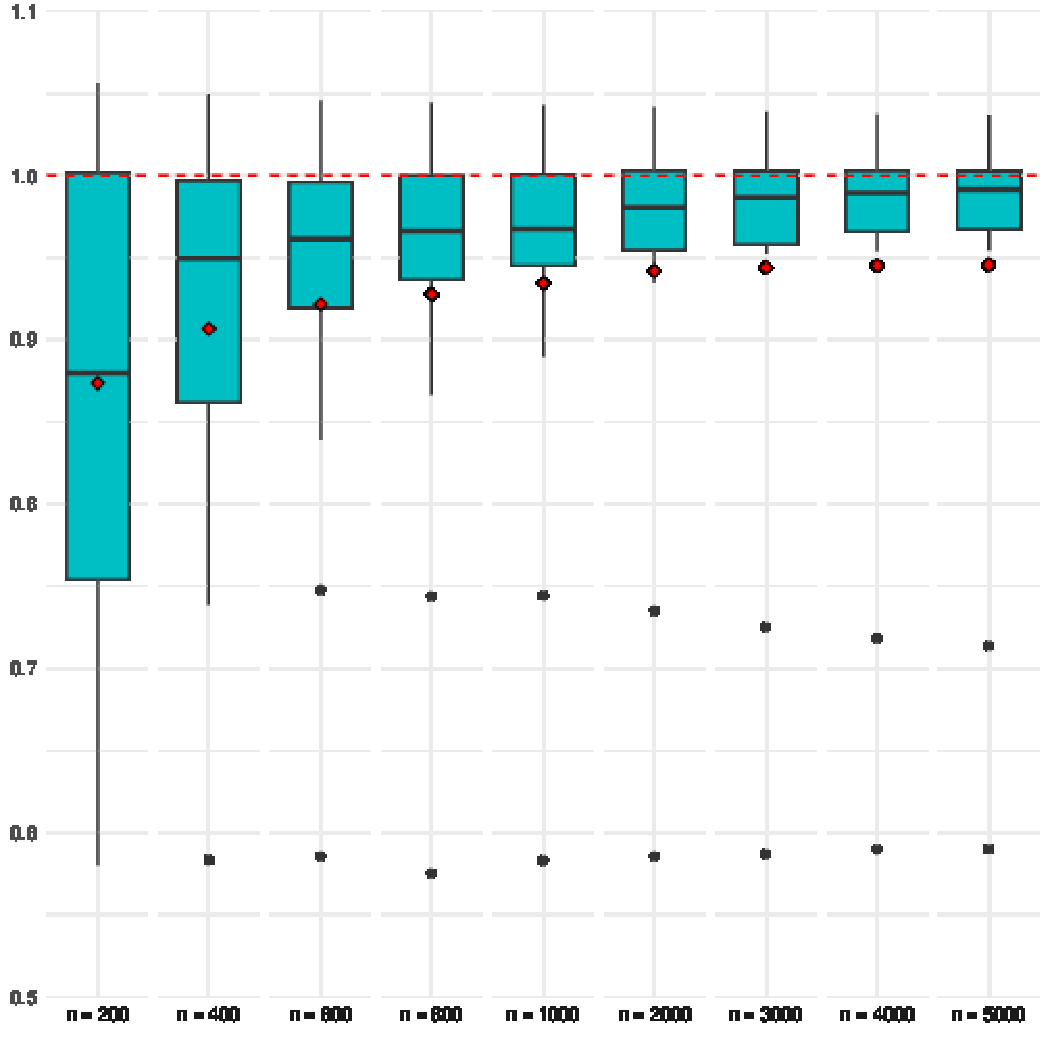}
\captionof{figure}{Boxplots of  RMSE ratios~\eqref{eq:rmse-rat}. For
  each training-set size $n$, the boxplot is based on the 14~ratios
  corresponding to the data sets listed in Table~\ref{table:datasets}.}
\label{fig:main}
\end{center}

\noindent
The results can be summarized as follows:
\begin{packed_itemize}
\item On balance, HRF clearly outperforms RF.
  
\item The gains are most pronounced for small training-set
  sizes $n$ but `live on'
  up to the largest size considered, $n=5000$.
  
\item Out of the 14 data sets, there is one on which HRF performs worse
  than RF, but the loss is never more than 6\% and always below 5\%
  for $n \ge 400$.

\item On the other hand, there are two data sets for which HRF reduces
  the RMSE compared to~RF
  by more than~25\% for $n \ge 400$; for one of these data sets,
  HRF actually reduces the RMSE by more than 40\% for all $n$.

\item Summing up, based on the 14 data sets considered, there is
  little to lose but potentially much to gain by upgrading from RF to HRF.  
  
  \item HRF outperforms RF particularly convincingly for smaller $n$,
    and thus for larger $d$ relative to~$n$.  The intuition for this
    finding is that both RF and HRF are `consistent'
  forecasting methods for sequences of data with $d$ independent of
  $n$ (when data are i.i.d.) and hence the difference between the
  two forecasts tend to decrease for larger $n$.

\end{packed_itemize} 


\begin{remark}[Cross-validation]
\label{rem:cv}
\rm   
At this point, some readers may wonder why we do not use
cross-validation to build up a $n \times p$ matrix of
pseudo-out-of-sample errors $R$. The reason lies in the special nature
of the random forest, since the individual forecasting methods, namely
the trees, depend on the underlying training set. If we used ten-fold
cross-validation, say, we would obtain ten different tree ensembles,
none of which would coincide with the ensemble used at the end for
forecasting, namely the ensemble based on the entire training set.
This problem would not arise with other forecasting methods, such as
linear regression models; of course, estimated parameters in
a given regression model would change
as a function of the underlying training set, but not the
`characteristics' of that model (such as the number
and the constitution of the regressors).
\end{remark}

\begin{remark}[Time series data]
\rm   
We have previously demonstrated that the hedged random forest, on
balance, provides superior forecasting performance compared to the
standard random forest (based on equally-weighted trees). To be fair,
all this evidence has been for cross-sectional (or i.i.d.) data sets.

We also experimented with time series data sets but failed to
outperform, on balance, the standard random forest. There could be
several reasons for this finding. On the one hand, with time series
data it is more difficult to estimate $\mu$ and $\Sigma$. On the other
hand, many real-life time series may not be stationary but suffer from
time-varying parameters or structural breaks; in such cases, the
estimates of $\mu$ and $\Sigma$ based on observations in a past window
(including today) could simply be noticeably off-target for what is
actually coming the future, and not weighting the trees would be more
robust.

Having said this, it may well be that for other applications
(different from the random forest) our generic high-level methodology
can also provide gains over simple averaging for time series data.
This topic
is left to future research.
\end{remark}  

\subsection{Related Literature}

Weighted versions of the random forest, that is, versions that do not
use equal weighting have, of~course, been considered before in the
literature.

One strand of this literature does not apply to our setting, since it
considers the random forest in the context of classification rather than
regression; for example, see \cite{kouloumpris:vlahavas:2022}.
In this case,
the goal is the minimize classification error rather than MSE which,
in principle, does not involve a covariance matrix and leads to
optimization problems of a different nature compared to those
considered in Section~\ref{sec:high-level}.

In the context of regression, the various proposals often are
problem-specific rather than generic, that is, they are designed
specifically for the random forest (only) rather than using a
high-level methodology as the one of Section~\ref{sec:high-level}
applied to the random forest as a special case (but not exploiting
specific features of it). Such proposals can be rather complex (to
understand) and difficult to implement (in terms of coding). As an
example, the reader is referred to \cite{chen:yu:zhang:2023}. It would
have been interesting to compare with their proposal
but both their Algorithms~1 and~2 are cumbersome to implement and the
authors do not provide (so far) any corresponding code. Furthermore,
their proposal requires the choice of tuning parameters which, in
addition to a training set, also requires a validation set,
resulting in a loss of information in practice;
see their Section~4.
On the other hand, we can compare to an earlier weighted random
forest which is quite easy to implement, namely the proposal of
\cite{winham:2013}. For each tree $\M_j$, they only consider the
corresponding out-of-bag (OOB) errors in the training set to compute
$$
t\mbox{PE}_j \defeq \frac{1}{|\text{OBB}_j|}
\sum_{i \in \text{OOB}_j} \bigl |y_i - \M_j(x_i)\bigr |~,
$$
where OOB$_j \subset \{1, \ldots, n\}$ denotes the OOB subset of the
training set corresponding to tree $\M_j$. Next, they compute
`relative' weights according to one of the three following formulas:
\begin{align}
\hat w_{j, \text{rel}} & \defeq 1 - t\mbox{PE}_j \label{eq:win1} \\
  \hat w_{j, \text{rel}} & \defeq \exp \left (\frac{1}{t\mbox{PE}_j} \right )
              \label{eq:win2}  \\
\hat w_{j, \text{rel}} & \defeq \left (\frac{1}{t\mbox{PE}_j} \right )^\lambda
            \quad \mbox{for some } \lambda > 0 \label{eq:win3}            
\end{align}
Finally, the weights $\{\hat w_j\}$ are forced to sum up to one by defining
$$
\hat w_j \defeq \frac{\hat w_{j, \text{rel}}}{\sum_{l=1}^p \hat w_{l, \text{rel}}}~.
$$
Note that, by construction, these weights $\hat w_j$ are all strictly positive.

Following the advice of \cite{winham:2013}, we tried as the
leading candidates version~\eqref{eq:win2} and 
version~\eqref{eq:win3} with $\lambda \defeq 5$, of which the former performed
somewhat better.
Calling the resulting method WRF (for Winham-et-al Random
Forest), we can construct as an analog to \eqref{eq:rmse-rat} the ratio

\begin{equation}\label{eq:rmse-rat-win}
\text{RMSE}_{\text{HRF}/\text{WRF}} \defeq 
  \frac{
  \sqrt{\frac{1}{B} \sum_{b=1}^B  \text{MSE}_{\text{HRF,b}}}}
  {\sqrt{\frac{1}{B} \sum_{b=1}^B  \text{MSE}_{\text{WRF,b}}}}~.
\end{equation}
As an analog to Figure~\ref{fig:main} we then obtain
Figure~\ref{fig:win}, which demonstrates that HRF clearly
outperforms~WRF.

\medskip
\begin{center}
\captionsetup{type=figure}
\includegraphics[width=5.5in,height=5in]
{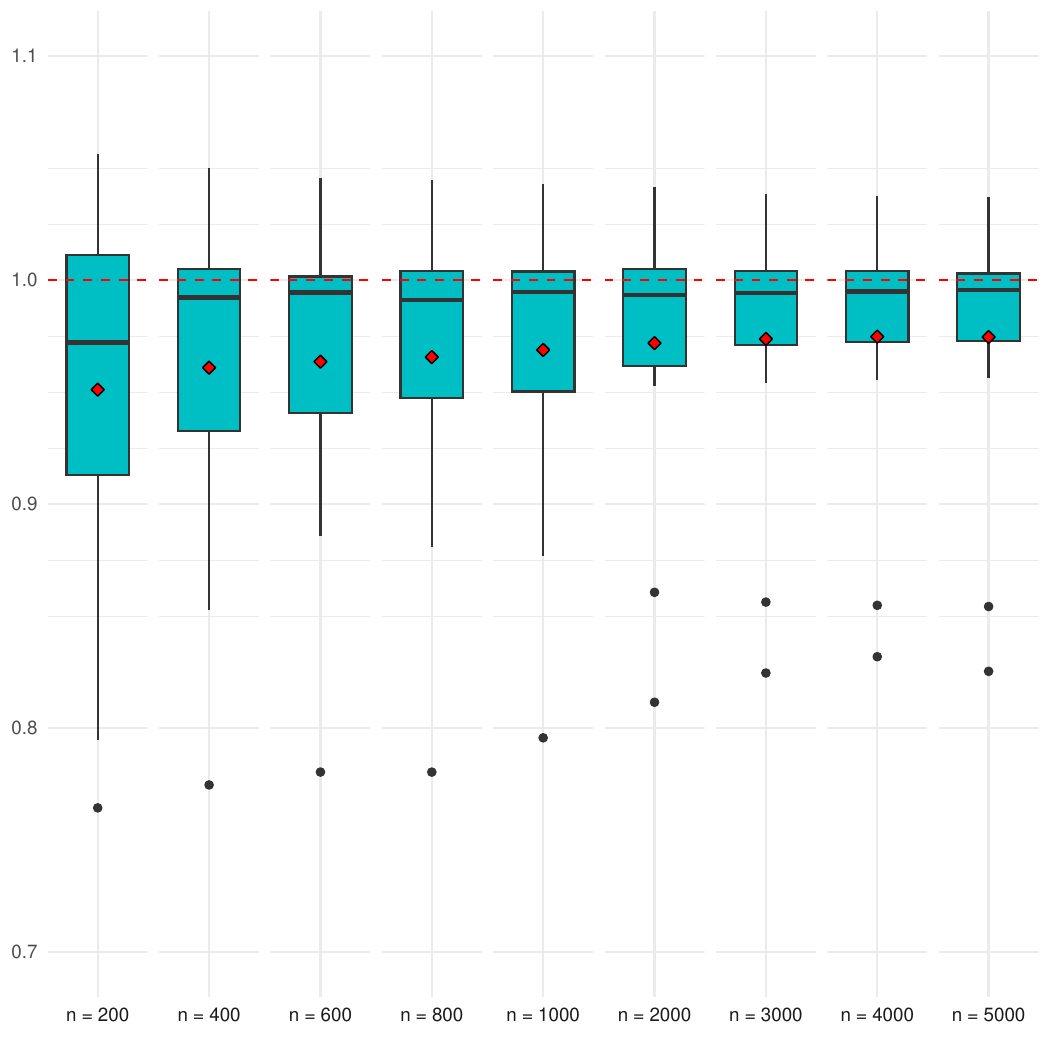}
\captionof{figure}{Boxplots of  RMSE ratios~\eqref{eq:rmse-rat-win}. For
  each training-set size $n$, the boxplot is based on the 14~ratios
  corresponding to the data sets listed in Table~\ref{table:datasets}.}
\label{fig:win}
\end{center}

\begin{remark}[Importance of negative weights]
  \label{rem:neg} \rm
Notably all previous proposals for weighting the random forest that
we are aware of, not only on the context of regression but 
also in the context of classification,
impose the ``no-short-sales constraint'' $\kappa = 1$, that is, $w_j
\ge 0 \; \forall j$. As shown in the robustness checks of
Appendix~\ref{app:robust},
allowing for negative
weights generally
improves performance and for certain data sets by a pronounced
margin. In the context of finance, a ``no-short-sales constraint'' can be
motivated by legislation (for example, mutual funds are not allowed to
short stocks) or by practical considerations (for example, shorting
certain assets may not be possible or prohibitively expensive). On the
other hand, ``short-selling'' individual forecast methods ${\cal M}_j$ by
assigning them a negative weight is always possible and
does not incur any monetary costs. We, therefore, hope that our paper will
serve as motivation to the scientific community to allow for negative
weights not only in the random forest but also 
in other forecast-combination applications.~\qed
\end{remark}


\newpage

\bibliographystyle{apalike}    
\bibliography{wolf}

\newpage

\begin{appendix}
\section{Robustness Checks}
  \label{app:robust}

The goal of this appendix is to provide robustness checks with respect
to the nature of the \mbox{estimator $\hat \Sigma$} and the choice of the
gross-exposure constraint $\kappa$.

As an alternative estimator $\hat \Sigma$ to nonlinear shrinkage,
we will consider the
canonical choice, the sample covariance matrix of the $n \times p$
matrix $R$. As far as $\kappa$ is concerned, we will consider the
choices $\kappa \in \{1, 1.5, 2, 2.5, \infty\}$., the canonical
choice being $\kappa \defeq 1$.

Figure~\ref{fig:all-kappa-nls} displays the results for all choices of
$\kappa$ when $\hat \Sigma$ is given by nonlinear shrinkage whereas
Figure~\ref{fig:all-kappa-sample} displays analogous results when
nonlinear shrinkage is replaced by the sample covariance matrix instead.

\begin{center}
\captionsetup{type=figure}
\includegraphics[width=5.5in,height=5in]
{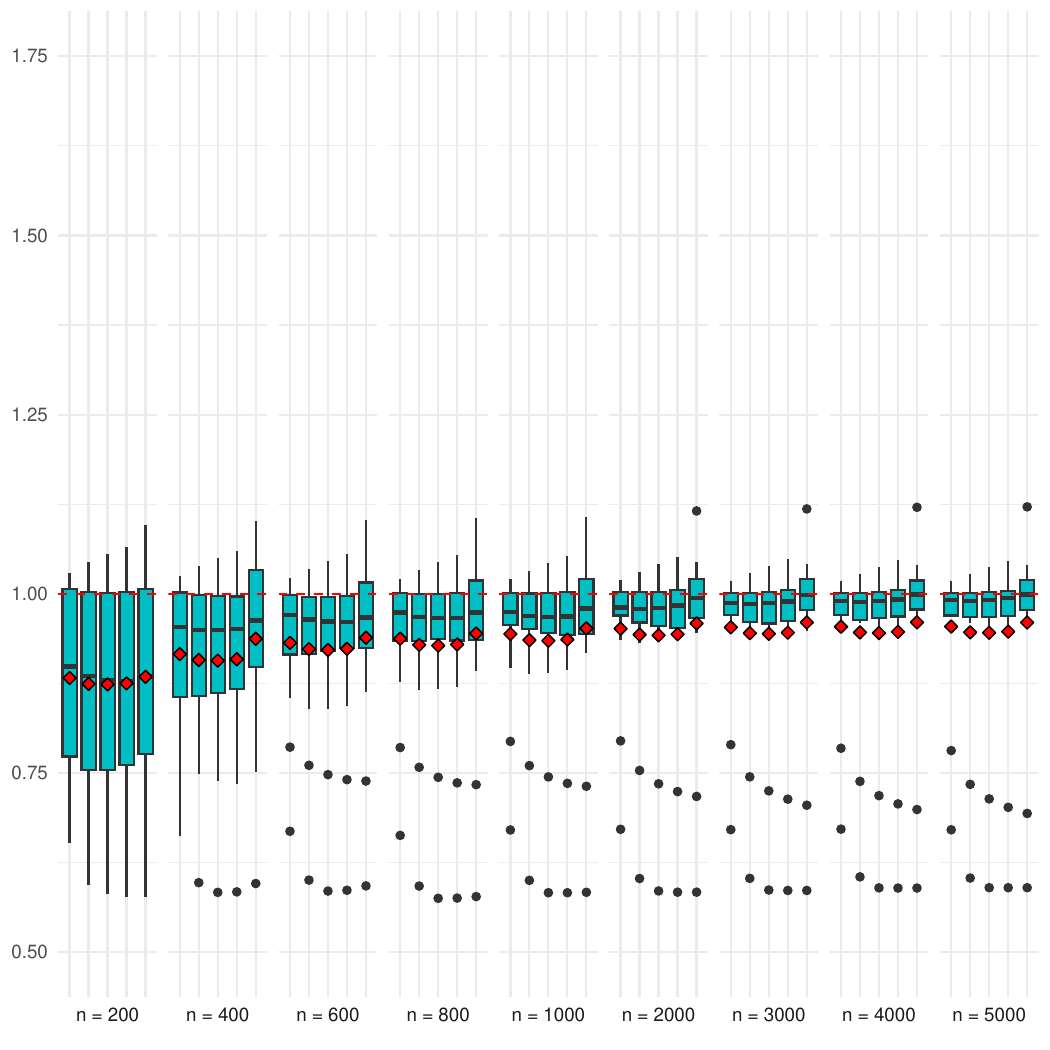}
\captionof{figure}{Boxplots of  RMSE ratios~\eqref{eq:rmse-rat}. For
  each training-set size $n$, there are five boxplots corresponding to
  $\kappa \in \{1, 1.5, 2, 2.5, \infty\}$ where
  each boxplot is based on the 14~ratios
  corresponding to the data sets listed in Table~\ref{table:datasets}.
  The estimator $\hat \Sigma$ is given by nonlinear shrinkage.}
\label{fig:all-kappa-nls}
\end{center}

\begin{center}
\captionsetup{type=figure}
\includegraphics[width=5.5in,height=5in]
{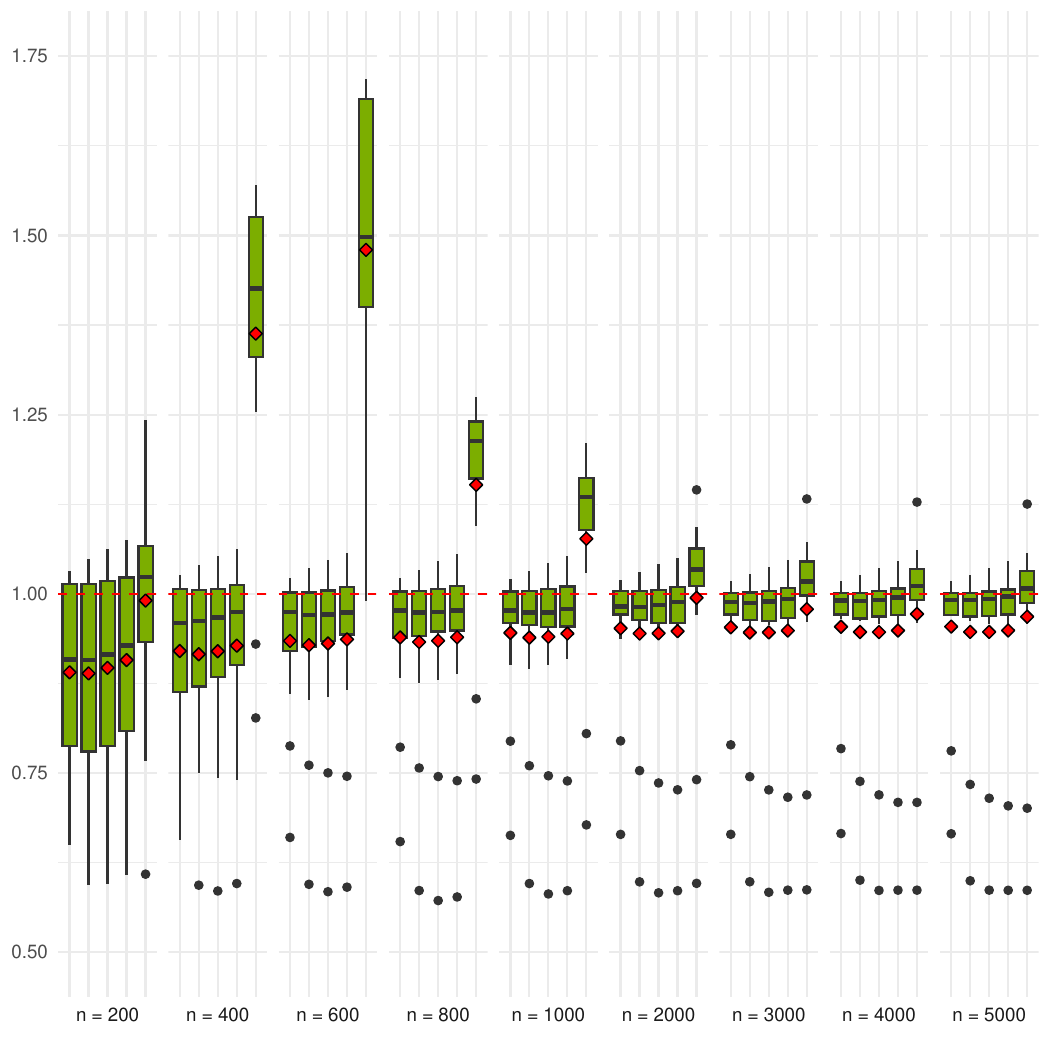}
\captionof{figure}{Boxplots of  RMSE ratios~\eqref{eq:rmse-rat}. For
  each training-set size $n$, there are five boxplots corresponding to
  $\kappa \in \{1, 1.5, 2, 2.5, \infty\}$ where
  each boxplot is based on the 14~ratios
  corresponding to the data sets listed in Table~\ref{table:datasets}.
  The estimator $\hat \Sigma$ is given by the sample covariance matrix.}
\label{fig:all-kappa-sample}
\end{center}

When $\hat \Sigma$ is given by nonlinear shrinkage, the choice $\kappa
\in \{1.5, 2, 2.5\}$ matters very little, and all three choices
perform better, in terms of both mean and median RMSE reduction,
compared to the canonical choice $\kappa \defeq 1$. Not imposing a
gross-exposure constraint (that is, the choice $\kappa \defeq \infty)$
performs worst overall, though it still outperform HR on balance.

When $\hat \Sigma$ is given by the sample covariance matrix instead,
the choice of $\kappa$ matters more. In particular, not imposing a
gross-exposure constraint (that is, the choice $\kappa \defeq \infty)$
can lead to grave underperformance of the HRF when the training-set
size $n$ is close to the number of trees $p$. Overall, the choice
$\kappa \defeq 1.5$ seems to perform best.

Finally, we look at the combined benefit of our version of HRF
compared to the canonical choice in the literature: (i) use nonlinear
shrinkage for $\hat \Sigma$ instead of the sample covariance matrix
and (ii) use $\kappa \defeq 2$ instead of $\kappa \defeq 1$.
Calling the two versions HRFour and HRFcan (for ``our'' and
``canonical''),
we can construct as an analog to \eqref{eq:rmse-rat} the ratio

\begin{equation}\label{eq:rmse-rat-can}
\text{RMSE}_{\text{HRFour}/\text{HRFcan}} \defeq 
  \frac{
  \sqrt{\frac{1}{B} \sum_{b=1}^B  \text{MSE}_{\text{HRFour,b}}}}
  {\sqrt{\frac{1}{B} \sum_{b=1}^B  \text{MSE}_{\text{HRFcan,b}}}}~.
\end{equation}

As an analog to Figure~\ref{fig:main} we then obtain
Figure~\ref{fig:can},
\medskip
\begin{center}
\captionsetup{type=figure}
\includegraphics[width=5.5in,height=5in]
{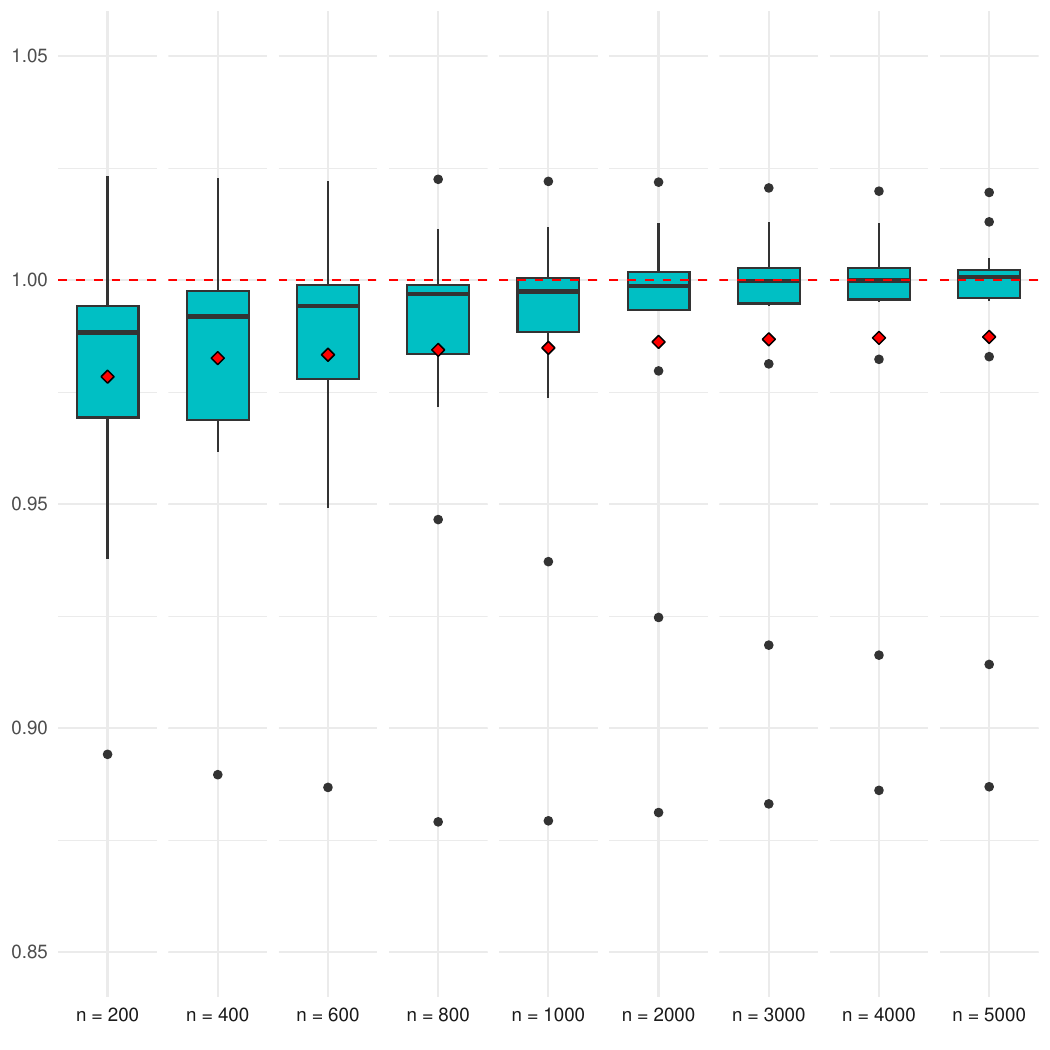}
\captionof{figure}{Boxplots of  RMSE ratios~\eqref{eq:rmse-rat-can}. For
  each training-set size $n$, the boxplot is based on the 14~ratios
  corresponding to the data sets listed in Table~\ref{table:datasets}.}
\label{fig:can}
\end{center}

One can see that in terms of the median RMSE ratio,
our version outperforms the
canonical version for $n \le 2000$, after which the competition becomes a tie,
whereas in terms of the mean RMSE ratio, 
our version outperforms the canonical version for all $n$.

\end{appendix}

\end{document}